\documentclass[%
reprint,
showpacs,
 amsmath,amssymb,
 aps,
 pra,
 10pt,
 twocolumn,
 floats,
]{revtex4-1}

\usepackage[latin1]{inputenc}
\usepackage{graphicx}
\usepackage{bm}
\usepackage{units}
\usepackage{SIunits}
\usepackage{color}

\newcommand{\mf}{m_F}
\newcommand{\Boff}{B_\mathrm{off}}
\newcommand{\Bfict}{B_\mathrm{fict}}
\newcommand{\vBoff}{\mathbf{B}_\mathrm{off}}
\newcommand{\vBfict}{\mathbf{B}_\mathrm{fict}}
\newcommand{\field}{\boldsymbol{\mathcal{E}}}
\newcommand{\ellipt}{\boldsymbol{\epsilon}}

\newcommand{\gauss}{\mathrm {G}}

\newcommand{\ket}[1]{\ensuremath{ \left|#1\right\rangle}}

\begin{document}


\title{Exploiting the local polarization of strongly confined light for sub-micron-resolution internal state preparation and manipulation of cold atoms}

\author{R. Mitsch}
\author{C. Sayrin}
\author{B. Albrecht}
\author{P. Schneeweiss}
\author{A. Rauschenbeutel}
\email{Arno.Rauschenbeutel@ati.ac.at}
\affiliation{%
 \mbox{Vienna Center for Quantum Science and Technology, Atominstitut, TU Wien, Stadionallee 2, 1020 Vienna, Austria}
}%


\begin{abstract}

A strongly confined light field necessarily exhibits a local polarization that varies on a subwavelength scale. We demonstrate that a single optical mode of such kind can be used to selectively and simultaneously manipulate atomic ensembles that are less than a micron away from each other and equally coupled to the light field. The technique is implemented with an optical nanofiber that provides an evanescent field interface between a strongly guided optical mode and two diametric linear arrays of cesium atoms. Using this single optical mode, the two atomic ensembles can simultaneously be optically pumped to opposite Zeeman states. Moreover, the state-dependent light shifts can be made locally distinct, thereby enabling an independent coherent manipulation of the two ensembles. Our results open the route towards advanced manipulation of atomic samples in nanoscale quantum optics systems.

\end{abstract}
\maketitle



\section{Introduction}
Over the past decades, experimentalists have gained full control over the internal and external degrees of freedom of individual quantum systems like single trapped ions~\cite{Wineland13} or atoms~\cite{Meschede06}. Coupling these quantum emitters to a single mode light field then allows one to, e.g., manipulate light at the quantum level~\cite{Haroche13}, interconnect distant quantum systems~\cite{Ritter2012}, and prepare entangled states of light and matter~\cite{Monroe04,Volz06}. Extending these techniques to ensembles of atoms or ions gave rise to a wealth of interesting phenomena and applications, in particular in the context of efficient light--matter coupling~\cite{Hammerer10} and the storage and retrieval of quantum states of light~\cite{Choi08,Lvovsky09,Jensen11}.

Recently, numerous experiments have made use of strongly confined light fields to advance the quantum manipulation of atomic samples, enabling, e.g., the enhancement of the coupling between atoms and light in nanophotonic devices~\cite{Vetsch10, Goban12, Thompson13b, Goban14} or the localization of atoms in optical microtraps~\cite{Schlosser01, Kaufman12, Thompson13}. In this article, we demonstrate that the unique polarization patterns of such tightly confined fields can be used to independently and simultaneously manipulate two distinct atomic ensembles with a single laser field in a way that is not possible with conventional light. We first show that one of two ensembles can be optically pumped to one Zeeman state while the other is simultaneously pumped to the opposite state. We then demonstrate that by using light-induced fictitious magnetic fields~\cite{Cohen-Tannoudji72} the ground state energy levels of the atoms can be shifted differently for the two ensembles~\cite{Lundblad09}, thereby enabling an independent coherent manipulation of each ensemble. This technique makes it possible for a single mode light field to interact with a medium consisting of two different classes of atoms and to thereby tailor the dispersive and absorptive properties of the medium independently~\cite{Proite08}. Moreover, it holds great promise for the realization of ultra-strong optical non-linearities down to the single photon level~\cite{Lukin00}.

\begin{figure}[tb]%
\includegraphics[width=0.95\columnwidth]{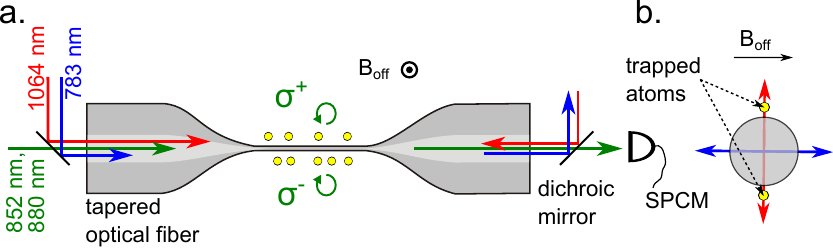}%
\caption{{\bf a.} Sketch of the experimental set-up including the tapered optical fiber, the laser fields, and the single-photon counting module (SPCM). The mostly circular polarization of the manipulation light fields (green line) at the position of the atoms (yellow dots) is indicated by the curved arrows and lies in the plane of the figure, denoted by $\mathcal{P}$. {\bf b.} Cross-sectional view of the nanofiber displaying the orientation of the linear polarizations of the blue- and red-detuned trapping fields, the atoms, and the magnetic offset field.
}%
\label{Fig:Setup}%
\end{figure}

Here, an optical nanofiber provides the strong confinement of the light fields~\cite{LeKien04b}. Nanofiber-based optical dipole traps, in which ensembles of a few thousand neutral atoms are trapped in the evanescent field surrounding the sub-wavelength-diameter silica fiber, have proven to provide an efficient interface between the nanofiber-guided light and the laser-cooled atoms~\cite{Vetsch10,Goban12}: High optical depths of a few percent per atom have been demonstrated and nanofiber-trapped atoms have been coherently manipulated with coherence times in the millisecond range~\cite{Reitz13}. This nanoscale quantum optics system is thus ideally suited for the implementation of the presented techniques.

\section{Experiment}
The experimental setup is depicted in Fig.~\ref{Fig:Setup}. Cesium atoms are trapped in the evanescent field surrounding an optical nanofiber (nominal radius $a = 250\,\nano\meter$), realized as the waist of a tapered optical fiber (TOF). A blue-detuned running wave with a free-space wavelength of $783\,\nano\meter$ and a power of $8.5\,\milli\watt$ and a red-detuned standing wave at $1064\,\nano\meter$ wavelength with $0.77\,\milli\watt$ per beam are sent through the TOF and create the trapping potential. Two diametric arrays of individual trapping sites are created $230\,\nano\meter$ above the nanofiber surface~\cite{Vetsch10}. The radial, azimuthal, and axial trap frequencies are $120$, $87$, $186\,\kilo\hertz$, respectively. Each trapping site contains at most one atom, and the average filling factor is $\lesssim 0.5$~\cite{Vetsch12}. These two individual one-dimensional arrays, separated by $\lesssim 1\,\micro\meter$, are a fraction of a millimeter long and correspond to two a priori equivalent atomic ensembles.

Any light field propagating in the optical nanofiber couples simultaneously to the two atomic ensembles. For all optical wavelengths involved in this experiment, the nanofiber only guides the fundamental  $\mathrm{HE}_{11}$ mode~\cite{Snyder83}. Linearly polarized light that is coupled into this mode exhibits a significant non-zero longitudinal component of its electric field. The latter is $\pi/2$-phase shifted with respect to the transverse components and its modulus and sign depend on the azimuthal position around the nanofiber~\cite{LeKien13c}. The ellipticity vector $\ellipt = i (\field\times\field^*)/|\field|^2$, where $\field$ is the positive-frequency envelope of the electric field, is thus position-dependent. In particular, $|\ellipt|$ is maximal along the direction of the transverse polarization of the light field and, for a wavelength of $\lambda=852$~nm and at a distance of 230~nm from the nanofiber surface, reaches about $0.84$. Moreover, $\ellipt$ has opposite signs on opposite sides of the fiber~\cite{Reitz14}. Hence, if the transverse polarization of the nanofiber-guided light lies in the plane $\mathcal{P}$ containing the trapped atoms (see Fig.~\ref{Fig:Setup}.a), $\field$ is almost fully $\sigma^+$ polarized above the nanofiber while it is almost fully $\sigma^-$ polarized below. Here, the quantization axis has been taken perpendicular to~$\mathcal{P}$.

\begin{figure}[t]%
\includegraphics[width=0.95\columnwidth]{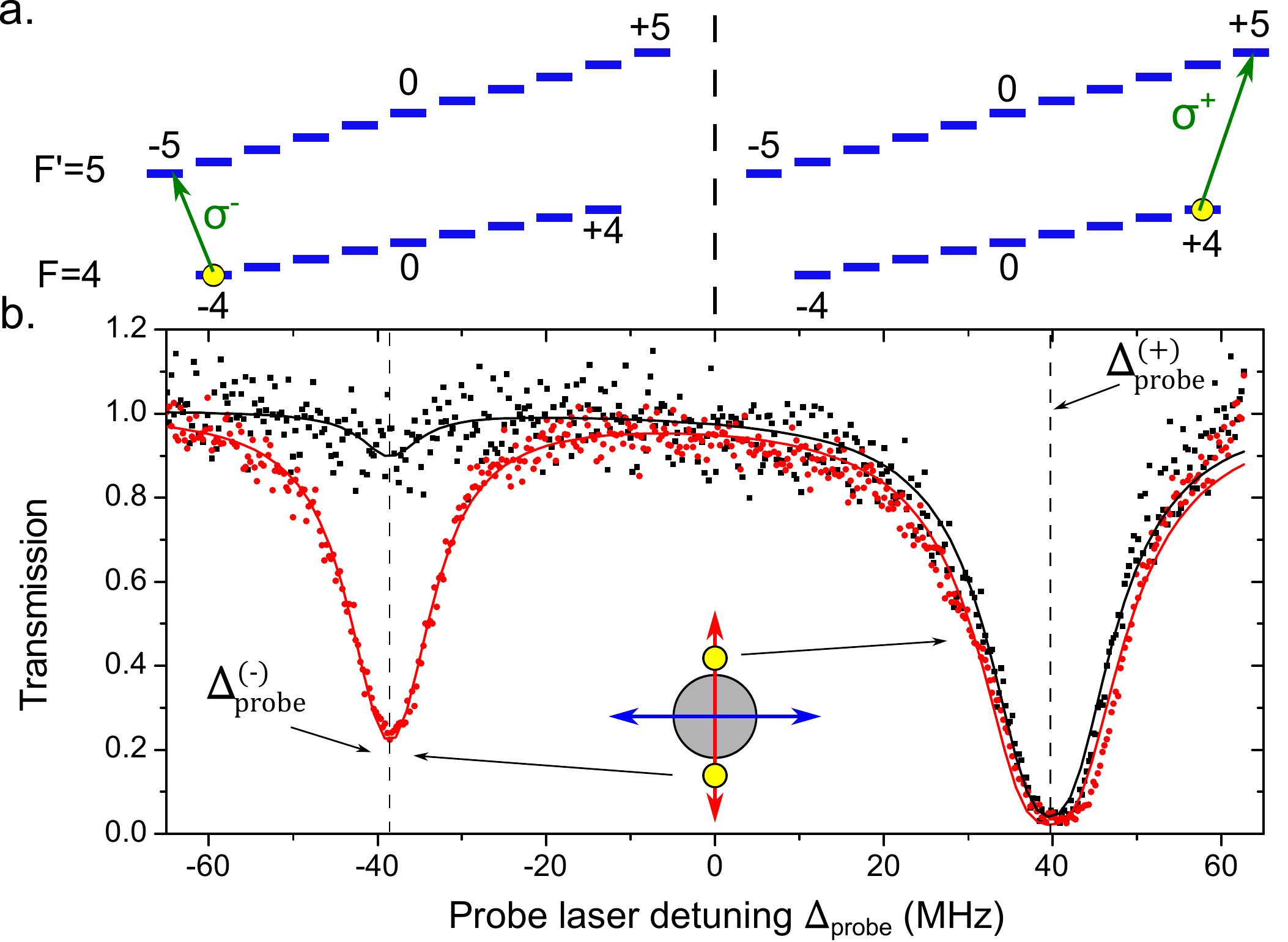}
\caption{%
{\bf a.} Schematic of the atomic energy levels. The yellow dots indicate the ideal population distribution of the two optically pumped atomic ensembles, that are trapped above (right panel) and below (left panel) the nanofiber. The arrows indicate, for each ensemble, the transition driven by the probe laser at resonance. {\bf b.} Transmission spectrum of the probe light field recorded for $\Boff=28\,\gauss$. The frequency is given relative to the measured resonance for $\Boff=0\,\gauss$. The two absorption dips at $\Delta_{\rm probe}^{(\pm)}$ stem from the two atomic ensembles, trapped above and below the fiber, as indicated by the yellow dots in the inset. The black (red) dots are recorded after (without) applying the push-out laser. The solid lines are fits to the data (see text).
}%
\label{Fig:Pumping}%
\end{figure}

\subsection{Optical Pumping}
We demonstrate that this polarization pattern can be used to optically pump the trapped atoms above the fiber into the outermost Zeeman sub-state $\ket{F=4, \mf=+4}$ while the atoms below the fiber are pumped to $\ket{F=4, \mf=-4}$, see Fig.~\ref{Fig:Pumping}.a. Once the atoms are loaded into the nanofiber-trap, they are in a mixture of $m_F$-states of the $F=4$ hyperfine manifold~\cite{Vetsch10}. We then apply a magnetic offset field $\vBoff$ of 28~G perpendicular to $\mathcal{P}$ and send a manipulation light pulse through the TOF. This magnetic field is large enough to clearly spectrally distinguish the two states and to prevent spin flips~\cite{Reitz13,LeKien13c}. The light pulse is resonant with the AC-Stark shifted $F=4\to F'=5$ transition of the Cs D2 line ($\lambda = 852\,\nano\meter$) and its transverse polarization lies in $\mathcal{P}$. The frequency of the manipulation light is scanned over $135\,\mega\hertz$ in $1\,\milli\second$ to address all $m_F$-states. The transmission spectrum  $T(\Delta_\text{probe})$ of a subsequent probe pulse is recorded within $5\,\milli\second$ by a single-photon counting module and plotted as red dots in Fig.~\ref{Fig:Pumping}.b. The power of the manipulation and probe laser fields is $4\,\pico\watt$. Two resonances are clearly observed as two dips in the transmission spectrum. The spectrum is well described by 
\begin{align}
\label{eq_satlorentz}
	T(\Delta_\text{probe})=\exp\left[ \sum_{k=+,-}-\frac{\text{OD}_k}{1+\frac{4\left(\Delta_\text{probe}-\Delta_\mathrm{probe}^{(k)}\right)^2}{\Gamma^2}} \right],
\end{align}
where the exponent accounts for the two Lorentzian line profiles corresponding to two optical transitions resonant at $\Delta_\mathrm{probe}^{(\pm)}$ (solid lines in Fig.~\ref{Fig:Pumping}.b). The Lorentzians are taken to have a common linewidth $\Gamma$ and individual optical densities (OD$_{\pm}$). We find a very good agreement for the detunings $\Delta_{\rm probe}^{(+)} = 39.82(8)\,\mega\hertz$ and $\Delta_{\rm probe}^{(-)} = -38.55(7)\,\mega\hertz$ and for $\Gamma=8.3(2)\,\mega\hertz$. The fitted linewidth is slightly larger than the $5.2\,\mega\hertz$ natural linewidth because of inhomogeneous light shifts induced by the trapping lasers. The frequency difference $\Delta_{\rm probe}^{(+)}-\Delta_{\rm probe}^{(-)}=78.4(1)\,\mega\hertz$ is in perfect agreement with the splitting between the outermost $\sigma^+$ and $\sigma^-$ $F=4\to F'=5$ transitions of 78.4~MHz for $\Boff=28\,\gauss$~\cite{Steck10}. This confirms that,  upon interaction with the manipulation field, the atoms above and below the fiber are optically pumped towards the \ket{F=4,\mf=+4} and \ket{F=4, \mf=-4} states, respectively. Finally, the fit yields the optical densities (OD$_{\pm}$) of the atomic medium at $\Delta_{\rm probe}^{(\pm)}$. Thus, by recording such a transmission spectrum of the resonant probe, we measure, within the same experimental run, the two ODs of the two atomic ensembles that are trapped above and below the fiber, respectively.

The demonstrated side-dependent optical pumping allows us to selectively address and, e.g., expel the atoms on only one side of the nanofiber from the trap. For this purpose, all atoms are exposed for $5\,\milli\second$ to a $\sigma^{-}$-polarized push-out laser beam~\cite{Kuhr05} that is resonant with the AC-Stark shifted $\ket{F=4, \mf=-4}\to\ket{F'=5, \mf=-5}$ transition and that propagates along $\vBoff$. The probe transmission spectrum is then measured as before, after an additional $1\,\milli\second$-long optical pumping sequence, see black squares in Fig.~\ref{Fig:Pumping}.b. While the dip at $\Delta_{\rm probe}^{(+)}$ has not been significantly affected by the push-out laser, the dip at $\Delta_{\rm probe}^{(-)}$ is now barely visible: The atoms trapped below the fiber are expelled from the trap while the atoms above the nanofiber are unaffected. This technique is particularly useful if one wants to couple trapped atoms to other quantum devices, such as SQUIDs~\cite{Hafezi12} or photonic structures: Knowing on which side of the fiber the atoms are prepared gives one full control over their position~\cite{schneeweiss13} with respect to the device. Furthermore, this technique paves the way to realize a directional nanophotonic atom--waveguide interface with our system~\cite{Mitsch14b}. Furthermore, this technique allows us to employ spin-orbit coupling of the nanofiber-guided light to direct the spontaneous emission of photons by an atom into a specific propagation direction~\cite{Mitsch14b}.

\subsection{Discerning and manipulating atoms via fictitious magnetic fields}
\begin{figure}[t]%
\includegraphics[width=0.95\columnwidth]{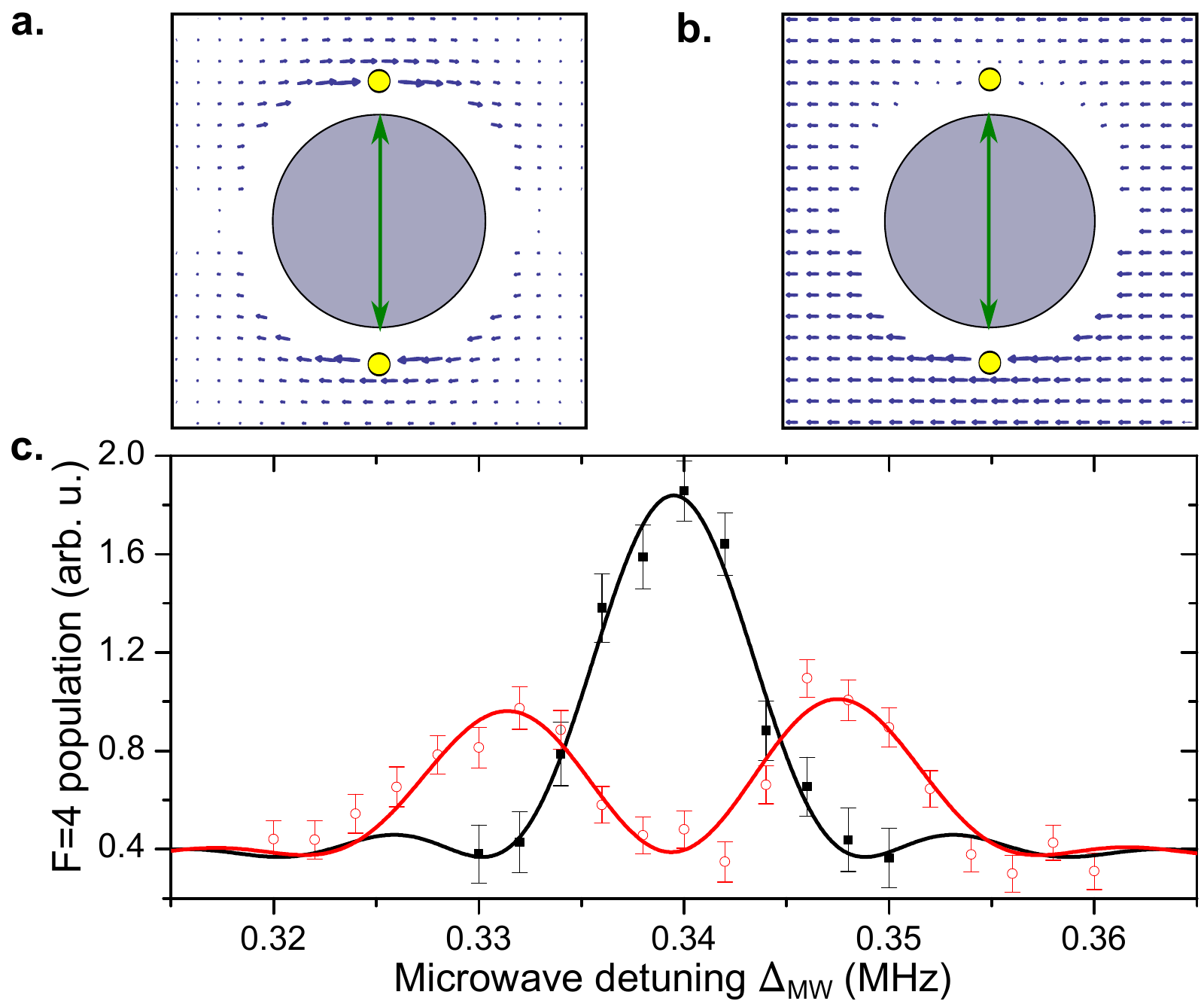}%
\caption{ {\bf a.}, {\bf b.} Cross-sectional view of the nanofiber indicating the position of the atoms (yellow dots) and the magnetic field $\vBoff+\vBfict$ (blue arrows), where $\vBfict$ is induced by a detuned quasi-linearly polarized light field (green double arrow): {\bf a.} $\Boff=0$, {\bf b.} $\Boff>\Bfict$ {\bf c.} Microwave spectrum of the $\ket{F=3, \mf=0} \to \ket{F=4, \mf=0}$ transition. The frequency is given relative to the zero-field transition frequency in free space. The black squares (red circles) are obtained with $\Bfict=0$ ($\Bfict \neq 0$). The solid lines are fits using a Fourier-limited line shape.
}%
\label{Fig:880}%
\end{figure}

We now show that it is possible to discern and to individually manipulate the two ensembles even when prepared in the same Zeeman substate. This would allow one, for example, to work with both ensembles in the least magnetic field-sensitive \ket{F,\mf=0} states in order to reach long coherence times~\cite{Reitz13}. For this purpose, we make the transition frequencies between the hyperfine ground states position-dependent, and thus different for the two atomic ensembles. This is achieved by using the $m_F$-state-dependent AC Stark shift induced by a far-detuned nanofiber-guided field: The AC Stark shift of the alkali atom ground states can be decomposed into an $m_F$-state-independent scalar and an $m_F$-state-dependent vector part~\cite{LeKien13}. The latter can be expressed as the effect of a fictitious magnetic field~\cite{Cohen-Tannoudji72} 
\begin{align}
\vBfict &= \beta^{(v)} i \field\times\field^* \, 
							= \, \beta^{(v)} |\field|^2 \ellipt,
\label{Eq:Bfict}
\end{align}
where $\beta^{(v)}$ is proportional to the vector polarizability of the Cs ground states~\cite{LeKien13}. 
Like for the ellipticity vector, $|\vBfict|$  is maximal along the direction of transverse polarization and $\vBfict$ has opposite signs on opposite sides of the fiber~\cite{LeKien13c}. Thus, in the presence of an additional detuned nanofiber-guided light field, the two atomic ensembles are subjected to opposite fictitious magnetic fields which are maximized if the polarization of this light field lies in $\mathcal{P}$ (see Fig.~\ref{Fig:880}.a). The linear Zeeman effect then leads to the desired lift of degeneracy between the two nanofiber sides for first order magnetic field sensitive transitions. In the case of transitions that only exhibit a quadratic Zeeman shift, like the $\ket{F=3, \mf=0}\to\ket{F=4, \mf=0}$ hyperfine clock transition, a non-zero external offset field $\vBoff$ must be applied in order to make the modulus of the total magnetic field side-dependent and thus to lift the degeneracy. In Fig.~\ref{Fig:880}.b, we depict this situation for $\vBoff \perp \mathcal{P}$ and $\Boff>\Bfict$. Above or below the nanofiber, at the positions of the two atomic ensembles, we then have $|\vBoff+\vBfict|=B_{\rm off}\pm B_{\rm fict}$, respectively.

While the light-induced fictitious magnetic field can be used to make the transition frequencies side-dependent, one has to avoid a significant distortion of the trapping potential by an unwanted scalar shift that is induced by the same laser: The strong radial intensity gradient of the nanofiber-guided fields~\cite{LeKien04b} results in a corresponding gradient of this scalar shift that leads to additional dipole forces. It is, however, possible to circumvent this problem by a proper choice of the wavelength of the laser that induces $\vBfict$. First, we can operate the laser at a tune-out wavelength~\cite{Arora11}, for which the scalar shift of the ground states vanishes. And second, $\vBfict$ can be induced by one of the trapping lasers. In the trapping configuration of Fig.~\ref{Fig:Setup}, the fictitious magnetic fields induced by the red and blue trapping fields vanish at the position of the trapping minima. However, a slight modification of the imbalance between the two counter-propagating red fields or a tilt of the blue polarization by a small angle from its initial position results in $\Bfict\neq 0$ because the polarization at the position of the atoms becomes elliptical~\cite{LeKien13c}. At the same time, this does not significantly modify the scalar shift and thus leaves the trapping potential essentially unchanged. 

\subsubsection{Manipulation light field at the tune-out wavelength}
We first characterize the effect of a nanofiber-guided manipulation light field at the Cs tune-out wavelength of $880.2524~\nano\meter$ on the clock transition of the two atomic arrays. We use a tunable microwave (MW) source at $9.2\,\giga\hertz$ that drives transitions between the hyperfine ground states. The black squares in Fig.~\ref{Fig:880}.c show the measured MW spectrum in the presence of a magnetic offset field $\Boff=28\,\gauss$ without applying the manipulation light field. For this measurement, the atoms are first prepared in the $F=3$ manifold. A 103-$\micro\second$ long MW $\pi$-pulse is then applied and transfers the atoms that are initially in \ket{F=3,\mf=0} to the \ket{F=4,\mf=0} state. Finally, the population of the $F=4$ manifold is measured as in~\cite{Reitz13} and plotted as a function of the MW detuning $\Delta_{\rm MW}$. A single Fourier limited peak is observed, see black line: in this configuration, the two atomic ensembles are equivalent and contribute equally to the measured spectrum.

The red circles in Fig.~\ref{Fig:880}.c show the MW spectrum under the same conditions when applying the manipulation light field with a power of $100\,\micro\watt$. A clear splitting of the resonance of $16.6(6)\,\kilo\hertz$ is observed which we identify as the difference of the quadratic Zeeman shifts of the two atomic ensembles. Here, we enhanced the splitting by working with $\Boff\gg\Bfict$. In this regime, the frequency shifts of the clock transition are approximately given by $\alpha_0 \, (\Boff^2 \pm 2 \Boff\cdot\Bfict)$, where $\alpha_0=0.427\,\kilo\hertz\per\gauss^2$~\cite{Steck10}, so that the splitting becomes linear in $\Bfict$ and is proportional to $\Boff$. Moreover, this implies that the two transition frequencies should split symmetrically with respect to the case without manipulation light field. This is in agreement with our experimental observations, see Fig.~\ref{Fig:880}.c. From the measured splitting in conjunction with the strength of the offset field, we find $\Bfict=0.35\,\gauss$ at the position of the atoms. Taking into account the distance of about 1~$\micro\meter$ between the atomic ensembles, this corresponds to a gradient of $\Bfict$ of 70~T/m, one order of magnitude larger than what was previously achieved with light induced fictitious magnetic fields~\cite{Lundblad09}. 

The above technique has the advantage that the splitting of the transition frequencies of the two atomic ensembles can be varied as fast as one can modulate the optical power of the manipulation light field. However, we experimentally observed that large splittings come at the expense of a broadening of the hyperfine transition. We attribute this fact to the strong gradient of $\Bfict$ (see Fig.~\ref{Fig:880}.a) which leads to position-dependent level-shifts and thus to a distortion of the trap. In particular, calculations show that the radial position of the trapping minima of the two hyperfine states are displaced in opposite directions. As a consequence, for $\mf=\pm3$ states, whose coherence times are an order of magnitude smaller than those of the $\mf=0$ states in our experiment, the MW spectrum is significantly altered when inducing frequency splittings that are large enough to be resolved.

\subsubsection{Discerning atoms via the blue-detuned trapping light field}
We now characterize the scheme that solely relies on tilting the transverse polarization of the blue trapping field by a small angle $\varphi_{\rm B}$ to induce the fictitious magnetic field at the position of the atoms. For $\varphi_\mathrm{B} = 0\degree$ the polarization of the blue trapping laser at the position of the atoms is purely linear and there is, consequently, no fictitious magnetic field. As soon as $\varphi_\mathrm{B} \neq 0\degree$ the light field can be decomposed into two parts: one with its main axis of polarization orthogonal to the plane $\mathcal{P}$ containing the atoms, that is therefore still linearly polarized at the position of the atoms, and another part with its main axis of polarization in $\mathcal{P}$. The latter is almost fully circularly polarized at the position of the atoms and its intensity is proportional to $\sin^2(\varphi_\mathrm{B})$. The non-zero local ellipticity of the electric field $\field$ induces the desired fictitious magnetic field $\vBfict$. Like for the light field at the tune-out wavelength, it points in opposite directions for the atomic ensembles above and below the fiber. In conjunction with a homogenous magnetic offset field $\vBoff$ perpendicular to $\mathcal{P}$, this leads to different values of the total magnetic field at the position of the atomic ensembles above and below the fiber and, therefore, to different Zeeman shifts of the MW transition frequencies.

\begin{figure}[t]
\includegraphics[width=0.95\columnwidth]{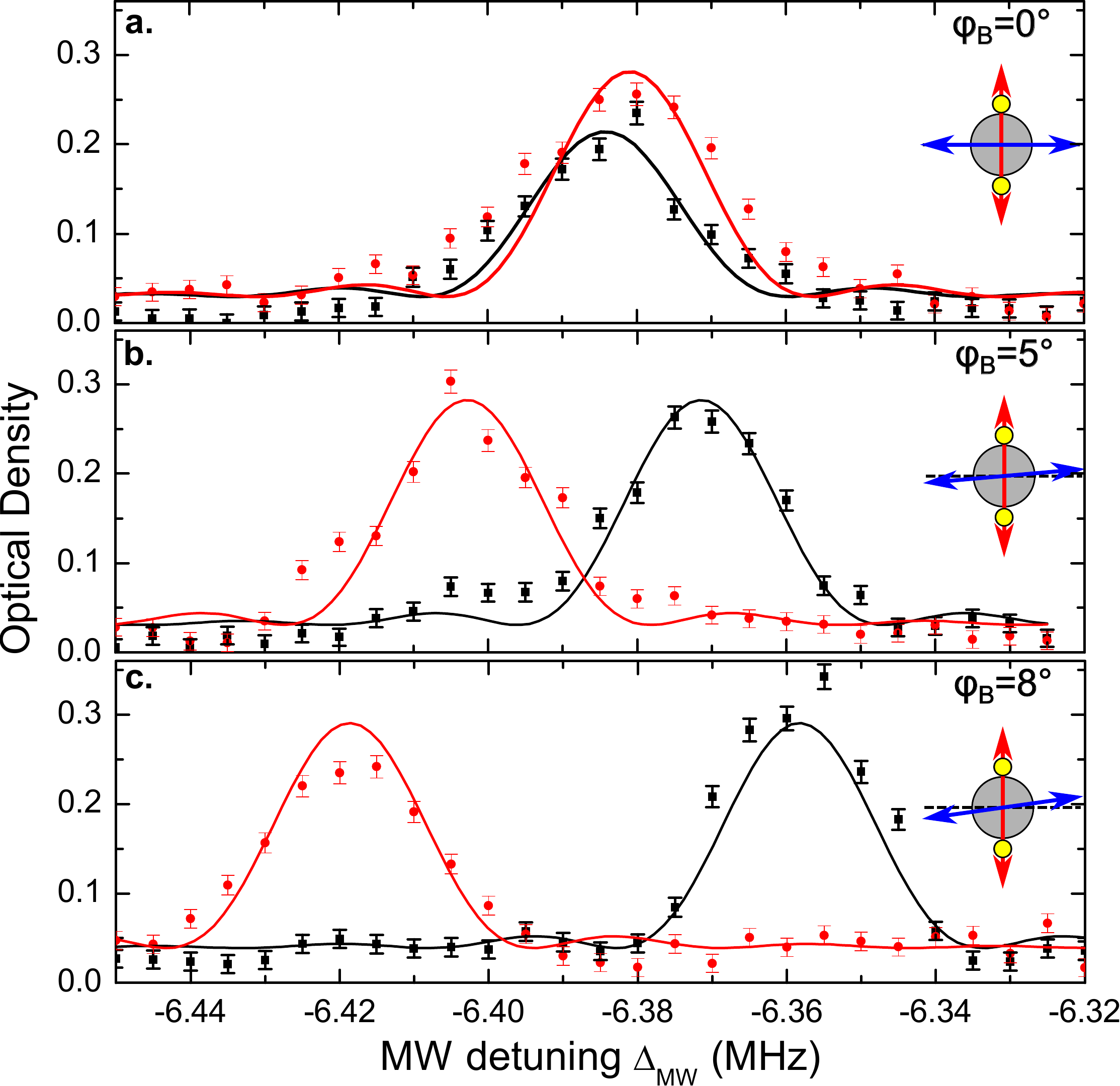}
\caption{Optical densities of the atomic ensembles trapped below (black squares) and above (red dots) the nanofiber, measured after a $40 \micro\second$-long MW pulse of frequency $\Delta_\mathrm{MW}$. The tilt angle, $\varphi_{\rm B}$, of the blue detuned trapping light field is indicated in the inset: {\bf a.} $\varphi_{\rm B}=0\degree$, {\bf b.} $\varphi_{\rm B}=5\degree$, {\bf c.} $\varphi_{\rm B}=8\degree$. The solid lines are fits using a Fourier limited line shape.}
\label{Fig:TiltedBlue}
\end{figure}

For this experiment we analyze the first-order magnetic-field sensitive $\ket{F=3, \mf=-3}\to \ket{F=4, \mf=-3}$ transition for three different values of $\varphi_\mathrm{B}$. We first load the Cs atoms in the $F=3$ hyperfine ground state. Here, we apply a magnetic offset field of $\Boff = 3\,\gauss$, for which two neighboring MW transitions are separated by $\approx1\,\mega\hertz$. A $40\,\micro\second$-long $\pi$-pulse at a given MW carrier detuning $\Delta_\mathrm{MW}$ then transfers the atoms that are initially in $\ket{F=3, \mf=-3}$ to $\ket{F=4, \mf=-3}$. After increasing the magnetic offset field to $28\,\gauss$, the atoms that were transferred to the $F=4$ manifold and are trapped above (below) the fiber are optically pumped to the $\ket{F=4, \mf=+4}$ ($\ket{F=4, \mf=-4}$) state. This is achieved, as described above, by sending a 1\,ms-long laser pulse, which is resonant with the $F = 4 \to F' = 5$ transition, with a wavelength of 852\,nm in the fiber. After the optical pumping sequence we record a probe transmission spectrum by sending the same laser light into the fiber and by scanning its frequency over the same interval within 5\,ms. The ODs of the two atomic ensembles are obtained by fitting the resulting transmission spectrum in the same way as discussed above (see Eq.~\eqref{eq_satlorentz}). Such transmission spectra have been recorded for different values of the MW detuning $\Delta_\mathrm{MW}$ and the obtained ODs are plotted in Fig.~\ref{Fig:TiltedBlue}.

This set of experiments was repeated for different values of the tilt angle $\varphi_\mathrm{B}$ of the blue trapping laser. For $\varphi_\mathrm{B}=0\degree$, the resonances for the two ensembles appear at the same MW frequency. For $\varphi_\mathrm{B}=5\degree$, however, the resonances are observed at different values of $\Delta_\mathrm{MW}$, corresponding to a splitting of 31.1(8)~kHz. This splitting further increases to $60.7(9)\,\kilo\hertz$ for $\varphi_\mathrm{B} = 8\degree$, without apparent inhomogeneous broadening of the line. This is almost four times larger than what was achieved with the manipulation light field at the tune-out wavelength. In particular, even with the first-order magnetic field sensitive transition used here, this allowed us to clearly spectrally discern the two atomic ensembles, see Fig.~\ref{Fig:TiltedBlue}.c.

Note that the frequency of the MW transition $\ket{F=3, \mf=-3} \to \ket{F=4, \mf=-3}$ is decreased (increased) for increasing values of $\varphi_\mathrm{B}$ for the atomic ensemble above (below) the fiber. This is consistent with the fact that the probe field and the blue trapping field are co-propagating and therefore have similar local polarizations at the position of the atoms, meaning that their ellipticity vectors  $\ellipt$ are parallel. Given that, for $\lambda=783\,\nano\meter$, the coefficient $\beta^{(v)}$ (Eq.~\eqref{Eq:Bfict}) is positive, the induced fictitious magnetic field is parallel to the ellipticity vector. Thus, above the fiber where the atoms are pumped to $\ket{F=4,m_F=+4}$, i.e., where $\ellipt$ is parallel to $\vBoff$, $\vBfict$ is as well parallel to $\vBoff$. The total magnetic field is $|\textbf{B}_\mathrm{total}|=\Boff+\Bfict$, leading to the fact that for the atoms above the fiber the frequency of the MW transition $\ket{F=3, \mf=-3} \to \ket{F=4, \mf=-3}$ is decreased. Below the fiber, where the atoms are pumped to $\ket{F=4,m_F=-4}$, i.e., where $\ellipt$ is anti-parallel to $\vBoff$, $\vBfict$ is anti-parallel to $\vBoff$. Therefore the total magnetic field below the fiber is $|\textbf{B}_\mathrm{total}|=\Boff-\Bfict$, leading to the fact that for the atoms below the fiber the frequency of the MW transition $\ket{F=3, \mf=-3} \to \ket{F=4, \mf=-3}$ is increased.

\section{Conclusion}
In summary, we showed that the sub-wavelength polarization patterns of nanofiber-guided light can be used to develop novel techniques for the quantum control of atoms. Side-dependent optical pumping was realized using a single optical mode and allowed us to simultaneously prepare two atomic ensembles in two different Zeeman sub-states, respectively. Moreover, light-induced fictitious magnetic fields were utilized to lift the degeneracy between the two atomic ensembles by a few 10~kHz, both for first order magnetic field sensitive and insensitive Zeeman sub-states. This made it possible to selectively address the two ensembles with MW radiation and to independently prepare them in different hyperfine states for any magnetic quantum number. Beyond the use of nanofiber-guided optical modes, our method can be implemented in many other quantum optical systems that make use of tightly confined light fields, like in optical tweezers~\cite{Schlosser01,Kaufman12,Thompson13}, plasmonic structures~\cite{Stehle11}, or nanophotonic devices~\cite{Thompson13b,Goban14}. Our results, thus, shed light on new opportunities offered by nanoscale quantum optics systems, and thereby provide new perspectives for future studies on quantum emitters coupled to nanophotonic devices~\cite{Chang08, Chang13, Griesser13}.

\section{Acknowledgments}
We acknowledge financial support by the Austrian Science Fund (FWF, SFB NextLite project No. F~4908-N23 and DK CoQuS project No. W~1210-N16). C.S. acknowledges support by the European Commission (Marie Curie IEF Grant 328545). R. M. and C. S. contributed equally to this work.


\bibliography{OneSide}

\end{document}